\documentclass[preprint,aps,%superscriptaddress,
showpacs%,nofootinbib,eqsecnum,preprintnumbers,twocolumn,prl
]{revtex4}

\usepackage{epsfig}

\input{epsf}

\newcommand{\be}{\begin{equation}}
\newcommand{\ee}{\end{equation}}
\newcommand{\bea}{\begin{eqnarray}}
\newcommand{\eea}{\end{eqnarray}}
\newcommand{\nn}{\nonumber \\}

\begin{document}

\preprint{Guchi-TP-018}
\date{\today%
}
\title{One-loop finite potential for $N-2$ scalars
from $N$ quantum fields}

\author{Yoshinori Cho}
\email{b2669@sty.cc.yamaguchi-u.ac.jp}
\affiliation{Graduate School of Science and Engineering, Yamaguchi
University,  Yoshida, Yamaguchi-shi, Yamaguchi 753-8512, Japan}

\author{Nahomi Kan}
\email{b1834@sty.cc.yamaguchi-u.ac.jp}
\affiliation{Graduate School of Science and Engineering, Yamaguchi University, 
Yoshida, Yamaguchi-shi, Yamaguchi 753-8512, Japan}

\author{Kenji Sakamoto}
\email{b1795@sty.cc.yamaguchi-u.ac.jp}
\affiliation{Graduate School of Science and Engineering, Yamaguchi
University,  Yoshida, Yamaguchi-shi, Yamaguchi 753-8512, Japan}

\author{Kiyoshi Shiraishi}
\email{shiraish@yamaguchi-u.ac.jp}
%\affiliation{Graduate School of Science and Engineering, Yamaguchi
%University,  Yoshida, Yamaguchi-shi, Yamaguchi 753-8512, Japan}
\affiliation{Faculty of Science, Yamaguchi University,
Yoshida, Yamaguchi-shi, Yamaguchi 753-8512, Japan}

\begin{abstract}

We study the one-loop effective potential induced
from quantum fluctuation of a finite number of fields. 
A series expansion in terms of the modified Bessel functions is
useful to evaluate the one-loop effective potential.
We find that at most $N-2$ scalars parameterize the one-loop finite
potential and  the explicit parameterization is shown.
The structure of the potential for $N=4$ is investigated
as the simplest case.
The implication of the model is discussed.

\end{abstract}



\maketitle

%%%%%%%%%%%%%%%%%%%%%%%%%%%%%%%%%%%%%%%%%%%%%%%%%%%%%%%%%%%%%%%%%%%%%%
%%%%%%%%%%%%%%%%%%%%%%%%%%%%%%%%%%%%%%%%%%%%%%%%%%%%%%%%%%%%%%%%%%%%%%
%%%%%%%%%%%%%%%%%%%%%%%%%%%%%%%%%%%%%%%%%%%%%%%%%%%%%%%%%%%%%%%%%%%%%%

%%%%%%%%%%%%%%%%%%%%%%%%%%%%%%%%%%%%%%%%%%%%%%%%%%%%%%%%%%%%%%%%%%%%%%
%%%%%%%%%%%%%%%%%%%%%%%%%%%%%%%%%%%%%%%%%%%%%%%%%%%%%%%%%%%%%%%%%%%%%%
\section{Introduction}
%%%%%%%%%%%%%%%%%%%%%%%%%%%%%%%%%%%%%%%%%%%%%%%%%%%%%%%%%%%%%%%%%%%%%%
%%%%%%%%%%%%%%%%%%%%%%%%%%%%%%%%%%%%%%%%%%%%%%%%%%%%%%%%%%%%%%%%%%%%%%

In general field theories, quadratic and logarithmic divergence appears
in the derivation of one-loop quantum corrections to some physical
parameters. We need a cut-off scale to regularize the loop integration.
This leads to a cut-off scale dependence of the one-loop potential and 
the source of the hierarchy problem in the unification theories.

In higher-dimensional theory, it is said that the one-loop finite
potential for extra-components of gauge field as scalars can be obtained
from (an infinite number of four-dimensional) quantum fields without
supersymmetry. The symmetry breaking mechanism according to such a
potential is called as the Hosotani mechanism~\cite{Hos}.
The reason why the finite potential is possible, in spite of the worse
degree of divergence in higher dimensions, is that the divergent part is
independent of the scalar degrees of freedom~\cite{ABQ,GNS}.

Recently a new type of theory,
which is known as deconstruction~\cite{ACG}, attracts much
attention. A number of copies of a four-dimensional theory
linked by a new set of fields can be viewed as a single theory.
The resulting theory may be almost equivalent to a 
higher-dimensional theory, but having a finite number of mass states.
It is pointed out that one-loop finite potential for a scalar
degree of freedom is obtained in deconstructing five-dimensional
QED~\cite{HL,KSS}. 

There may be an {\it inverse problem} : If we evaluate the one-loop
effect of
$N$ quantum fields, how many degrees of scalars can we have, in order
that their potential is finite?
We will show that the number can be obtained easily, and further,
we will give the parameterization of masses by the scalars explicitly.

%%%%%%%%

In this paper, we examine the $N$-scalar theory without
self-interactions, while the same technique is valid for the one-loop
effect of fermion fields. We parameterize the masses by scalar degrees of
freedom to analyze the effective potential. 
In Sec.~\ref{sec:2}, the mass spectrum of fields are parameterized
appropriately. The one-loop quantum effect of scalar fields with this
mass spectrum is calculated in Sec.~\ref{sec:3}. 
In Sec.~\ref{sec:4}, the free energy density is estimated and the
superficial dimensionality is argued. The simplest model for $N=4$ is
studied in Sec.~\ref{sec:5}, where the explicit structure of the
potential is shown.  We close with Sec.~\ref{sec:f}, where
summary and conclusion are given.

%%%%%%%%%%%%%%%%%%%%%%%%%%%%%%%%%%%%%%%%%%%%%%%%%%%%%%%%%%%%%%%%%%%%%%
%%%%%%%%%%%%%%%%%%%%%%%%%%%%%%%%%%%%%%%%%%%%%%%%%%%%%%%%%%%%%%%%%%%%%%
\section{parameterization of masses}
\label{sec:2}
%%%%%%%%%%%%%%%%%%%%%%%%%%%%%%%%%%%%%%%%%%%%%%%%%%%%%%%%%%%%%%%%%%%%%%
%%%%%%%%%%%%%%%%%%%%%%%%%%%%%%%%%%%%%%%%%%%%%%%%%%%%%%%%%%%%%%%%%%%%%%

Suppose $N$ real scalar fields without self-interactions. We assume
$N\ge 3$. Their (mass)${}^2$ eigenvalues are denoted as $M^2_p$ ($p=1, 2,
\ldots, N$). If these masses depend on scalars, the one-loop vacuum
energy would become the potential for the scalars.

We parameterize $M^2_p$ as
\be
M^2_p=\overline{M}^2-\sum_{r=1}^{[N/2]}a_r \cos\left(\frac{2\pi
r}{N}p\right) -\sum_{r=1}^{[(N-1)/2]}b_r\sin\left(\frac{2\pi
r}{N}p\right)\, ,
\ee
where $[z]$ is the largest integer which does not exceed $z$.
Of course, $N$ parameters $\overline{M}^2$, $a_r$ and $b_r$
are directly derived as
\be
\overline{M}^2=\frac{1}{N}\sum_{p=1}^{N}M^2_p \, ,
\ee
\be
a_r=-\frac{2}{N}\sum_{p=1}^{N}M^2_p 
\cos\left(\frac{2\pi r}{N}p\right)\, ,
\quad
b_r=-\frac{2}{N}\sum_{p=1}^{N}M^2_p\sin\left(\frac{2\pi
r}{N}p\right)\qquad {\rm for}~r=1,\ldots,\left[\frac{N-1}{2}\right]\, ,
\ee
and in addition for even $N$,
\be
a_{N/2}=-\frac{1}{N}\sum_{p=1}^{N}(-1)^p M^2_p \, .
\ee

For later use, we once rewrite $M^2_p$ as
\be
M^2_p=\overline{M}^2-
\sum_{r=1}^{[N/2]}f_r\cos\left(\frac{2\pi r}{N}p+\varphi_r\right)\, ,
\label{expa}
\ee
where 
$f_r=\sqrt{a_r^2+b_r^2}$ and $\varphi_r=-\arctan(b_r/a_r)$ for
$r=1,\ldots,[(N-1)/2]$, and for even $N$, $f_{N/2}\equiv a_{N/2}$ and
$\varphi_{N/2}\equiv 0$.

Furthermore, we can enlarge the region of $\varphi_r$ to $[0, 2\pi)$,
while the form (\ref{expa}) is unchanged. Now $N$ independent variables
which parameterize $M_p^2$ are
$\overline{M}^2$,
$f_r$ and
$\varphi_r$.

%%%%%%%%%%%%%%%%%%%%%%%%%%%%%%%%%%%%%%%%%%%%%%%%%%%%%%%%%%%%%%%%%%%%%%
%%%%%%%%%%%%%%%%%%%%%%%%%%%%%%%%%%%%%%%%%%%%%%%%%%%%%%%%%%%%%%%%%%%%%%
\section{the effective potential}
\label{sec:3}
%%%%%%%%%%%%%%%%%%%%%%%%%%%%%%%%%%%%%%%%%%%%%%%%%%%%%%%%%%%%%%%%%%%%%%
%%%%%%%%%%%%%%%%%%%%%%%%%%%%%%%%%%%%%%%%%%%%%%%%%%%%%%%%%%%%%%%%%%%%%%

In this section, we evaluate the quantum effect of the scalar fields
with masses (\ref{expa}).
The one-loop effective potential is obtained by
\bea
& &\lim_{D\rightarrow 4^{-}}\frac{-\mu^{4-D}}{2(2\pi)^{D}}\sum_{p}
\int_0^{\infty}\frac{dt}{t}~
\int d^{D}k~\exp\left[-(k^2+M_{p}^2)t\right]\nn
&=&\lim_{D\rightarrow
4^{-}}\frac{-\mu^{4-D}}{2(4\pi)^{D/2}}\int_0^{\infty}
\frac{dt}{t}t^{-D/2}~
\sum_{p}\exp\left[-M_{p}^2t\right]\, ,
\label{sch}
\eea
after the dimensional regularization.
Here $\mu$ has the dimension of mass.
We ignore $\mu$ in the following discussion, because the finite
potential will be independent of $\mu$. 

If we expand the exponential in (\ref{sch}) with respect to $t$,
we find that apparent divergences are proportional to
$\sum_pM_{p}^2$ and $\sum_pM_{p}^4$.
Thus we have two constraint $\sum_pM_{p}^2=const.$ and
$\sum_pM_{p}^4=const.$ on the scalar parameters
for the one-loop finiteness.
Therefore the maximum number of scalars is $N-2$
for the finiteness of potential.

Let us clarify the structure of the scalar potential and
how the potential is parameterized by $N-2$ scalars.
At this point, the following mathematical formula
is very useful~\cite{KSS,SSK,KS};
\be
\exp\left[t(\cos\theta)\right]=\sum_{\ell=-\infty}^{\infty}
\cos \ell\theta~ I_{\ell}(t)
=\sum_{\ell=-\infty}^{\infty}
e^{i\ell\theta} I_{\ell}(t)\, ,
\label{formu}
\ee
where $I_{\ell}(x)$ is the modified Bessel function,
which satisfies $I_{-\ell}(x)=I_{\ell}(x)$ for integer $\ell$~\cite{GR}.

{}From (\ref{expa}), (\ref{sch}) and (\ref{formu}), we find
\be
\sum_{p=1}^N\exp\left[-M_{p}^2t\right]=e^{-\overline{M}^2t}
\sum_{p=1}^N\prod_{r=1}^{[N/2]}\left[\sum_{\ell_r=-\infty}^{\infty}\exp
(i\ell_r(\frac{2\pi
rp}{N}+\varphi_r))I_{\ell_r}\left(f_rt\right)\right]\, .
\ee
Resumming the phases, we get
\be
\sum_{p=1}^N\exp\left[-M_{p}^2t\right]=e^{-\overline{M}^2t}
\sum_{\{\ell_r\}}\left\{
\left[\sum_{p=1}^N\exp
(i\frac{2\pi p}{N}\sum_{r=1}^{[\frac{N}{2}]}r\ell_r)\right]
\exp(i\sum_{r=1}^{[\frac{N-1}{2}]}\ell_r\varphi_r)
\prod_{r=1}^{[\frac{N}{2}]}I_{\ell_r}\left(f_rt\right)\right\}\, .
\ee
Carrying out the summation over $p$ first,
we find 
\be
\sum_{p=1}^N\exp\left[-M_{p}^2t\right]=N e^{-\overline{M}^2t}
{\sum_{\{\ell_r\}}}'\left[
\exp(i\sum_{r=1}^{[\frac{N-1}{2}]}\ell_r\varphi_r)
\prod_{r=1}^{[\frac{N}{2}]}I_{\ell_r}\left(f_rt\right)\right]\, ,
\ee
where
${\sum_{\{\ell_r\}}}'$ means that the summation is performed
over $\{\ell_r\}$ which satisfy
$\sum_{r=1}^{[\frac{N}{2}]}r\ell_r\equiv 0~(mod~N)$.

Since $I_{\ell}(x)\propto \exp{x}$ for large $x$,
the sufficient condition on the convergence of the integration
(\ref{sch}) at large
$t$ is
\be
\overline{M}^2-\sum_{r=1}^{[\frac{N}{2}]}|f_r|\ge 0\, .
\ee
This is just the positivity of all $M_p^2$.%
\footnote{If the inequality is not satisfied, the possible region of the
parameter has to be reduced.}

Let us examine the convergence of the integration (\ref{sch}) at small
$t$. Since $I_{\ell}(x)\approx (x/2)^{\ell}/\ell!$ for small $x$,
the divergences appear only the terms with 
$\sum_r|\ell_r|-3<0$.

For $N$ odd,
this holds only for all $\ell_r\equiv 0$,
because $\sum_{r=1}^{[\frac{N}{2}]}r\ell_r\equiv 0~(mod~N)$
is satisfied. 
Thus the divergence appears in the part
\be
-\frac{N}{2(4\pi)^{D/2}}\int_0^{\infty}
\frac{dt}{t}t^{-D/2}~
e^{-\overline{M}^2t}
\prod_{r=1}^{[\frac{N}{2}]}I_{0}\left(f_rt\right)\, .
\ee
Note that this part is 
independent of $\varphi_r$.

Since $I_{0}(x)\approx 1+x^2/4+\cdots$ for small $x$,
the divergence in the part behaves as
\be
-\frac{N\overline{M}^{D}}{2(4\pi)^{D/2}}\left[
\Gamma\left(-\frac{D}{2}\right)+\frac{1}{4}
\frac{\sum_{r=1}^{[N/2]}f_r^2}{\overline{M}^4}
\Gamma\left(2-\frac{D}{2}\right)\right]\, .
\label{odd}
\ee
These terms must be independent of scalars whose potential will be
finite. Thus
$\overline{M}^{2}$ and
$\sum_{r=1}^{[N/2]}f_r^2$ are required to be constant.

For $N$ even, other
divergences appear in the two terms with $\ell_{N/2}=\pm 2$ and
$\ell_r=0$ ($r=1,\ldots,(N-2)/2$). Note that this part is also
independent of $\varphi_r$.
The divergent contribution of the
terms are
\be
-\frac{2N\overline{M}^{D}}{2(4\pi)^{D/2}}\left[\frac{1}{8}
\frac{f_{N/2}^2}{\overline{M}^4}
\Gamma\left(2-\frac{D}{2}\right)\right]\, .
\label{even}
\ee
Combining (\ref{odd}) and (\ref{even}), $\overline{M}^{2}$ and
$\sum_{r=1}^{(N-2)/2}f_r^2+2f_{N/2}^2$ are required to be constant
if $N$ is even.

Defining $\bar{f}_r=f_r$ for $r=1,\ldots,[(N-1)/2]$ and
$\bar{f}_{N/2}=\sqrt{2}f_{N/2}$,
we simply state
that scalars
$\varphi_r$ and $\bar{f}_r$, which satisfies
$\sum_{r=1}^{[N/2]}\bar{f}_r^2=const.$, parameterize the one-loop finite
potential. The degree of freedom is $N-2$, as expected;
we remark
$\sum_pM^4_p=\overline{M}^4_p+\frac{1}{2}\sum_{r=1}^{[N/2]}\bar{f}_r^2$. 
The constraint on scalars should be expressed by a non-linear sigma model
with (at least locally)
$O([N/2])\otimes U(1)^{[(N-1)/2]}$  symmetric kinetic
term. The realization of the sigma model can be attained by the other
potential which leads to the vacuum expactation value of
$\sum_{r=1}^{[N/2]}\bar{f}_r^2$, as in an interpretation of
deconstructed models~\cite{HL}.

%%%%%%%%%%%%%%%%%%%%%%%%%%%%%%%%%%%%%%%%%%%%%%%%%%%%%%%%%%%%%%%%%%%%%%
%%%%%%%%%%%%%%%%%%%%%%%%%%%%%%%%%%%%%%%%%%%%%%%%%%%%%%%%%%%%%%%%%%%%%%
\section{the `apparent dimension' of spacetime}
\label{sec:4}
%%%%%%%%%%%%%%%%%%%%%%%%%%%%%%%%%%%%%%%%%%%%%%%%%%%%%%%%%%%%%%%%%%%%%%
%%%%%%%%%%%%%%%%%%%%%%%%%%%%%%%%%%%%%%%%%%%%%%%%%%%%%%%%%%%%%%%%%%%%%%

In some models of deconstruction, the limit of large number of fields
yields higher-dimensional theory~\cite{HL,KSS,SSK,Lane}.
Our model in the present paper can be regarded as a generalization of
deconstruction in some meaning. Is it possible that the spacetime looks
like higher dimensions in our model?

To see this, we calculate the free energy at finite temperature.
This is because the exponent of the leading term in the high temperature
expansion depends on the dimension of the spacetime. 
We define the `apparent dimension' as
\be
\lim_{T\rightarrow\infty}\frac{\partial\ln|F|}{\partial\ln T}\, ,
\ee
where $F$ is the free energy (density) and $T$ is the temperature.

To obtain the free energy, we replace the integration over the frequency
by the summation over the discrete Matsubara frequencies (and attach a
certain factor)\cite{ft}. The free energy density is then obtained by
\be
F=-\frac{1}{(4\pi)^{2}}\int_0^{\infty}
\frac{dt}{t}t^{-2}~
\sum_{p}\sum_{n=1}^{\infty}
\exp\left[-M_{p}^2t-\frac{\beta^2n^2}{4t}\right]\, ,
\ee
where $\beta=T^{-1}$.

The dominant dependence on temperature can be found in the part
\bea
& &-\frac{N}{(4\pi)^{2}}\sum_{n=1}^{\infty}\int_0^{\infty}\frac{dt}{t^3}~
\exp\left[-\overline{M}^2t-\frac{\beta^2n^2}{4t}\right] 
\prod_{r=1}^{[\frac{N}{2}]}I_{0}\left(f_rt\right)\, ,
\nn
&=&-\frac{N}{(4\pi)^{2}}\frac{1}{\beta^4}
\sum_{n=1}^{\infty}\frac{1}{n^4}
\int_0^{\infty}\frac{dt}{t^3}~
\exp\left[-\beta^2\overline{M}^2n^2t-\frac{1}{4t}\right]
\prod_{r=1}^{[\frac{N}{2}]}I_{0}(f_r\beta^2n^2t)\, ,
\label{oo}
\eea
We assume that $N$ is an sufficiently large number. Relabeling $\{f_r\}$
so that
$f_1\ge f_2\ge\cdots\ge f_{[N/2]}$, we assume the simplest situation,
$f_{c+1}\ll T\ll f_{c}$. Eq.~(\ref{oo}) can be approximated, with the
limiting form 
$I_{0}(z)\sim 1$ for a small argument
using $I_{0}(z)\sim e^z/\sqrt{2\pi
z}$ for a large argument, as
\be
-\frac{N}{(4\pi)^{2+c/2}}\frac{1}{\beta^{4+c}\prod_{i=1}^{c}
\sqrt{f_i/2}}
\sum_{n=1}^{\infty}\frac{1}{n^{4+c}}
\int_0^{\infty}\frac{dt}{t^{3+c/2}}~
\exp\left[-\beta^2(\overline{M}^2-\sum_{i=1}^c
f_i)n^2t-\frac{1}{4t}\right]
\, ,
\ee
For further simplicity, $\beta^2(\overline{M}^2-\sum_{i=1}^c
f_i)$ is assumed small. 
This condition garantees the sufficiently spreaded mass spectrum.
Then the leading behavior of the free energy
density turns out to be $F\propto -T^{4+c}$.
Thus the spacetime dimension seems $\approx 4+c$.
We can say that the volume of the `apparent extra space' is roughly given
as $\frac{N}{\prod_{i=1}^{c}\sqrt{f_i/2}}$, by looking the overall
factor of $F$.

The maximum number of `apparent dimension' is approximately $[N/2]$.
This fact is understood in terms of configuration of theory space.
Imagine the $[N/2]$ orthogonal axes. Take two points on each axis.
Associate a field theory with each point and lay link fields
between the fields in the different axes.
The minimal theory space configuration is roughly the configuration of
the vertex of an
$[N/2]$-dimensional generalization of an octahedron.

Note that the present estimation is very rough.
If $N$ is small, the free energy is very sensitive to temperature $T$ and
the superficial dimensionality given here does not make any sense.

%%%%%%%%%%%%%%%%%%%%%%%%%%%%%%%%%%%%%%%%%%%%%%%%%%%%%%%%%%%%%%%%%%%%%%
%%%%%%%%%%%%%%%%%%%%%%%%%%%%%%%%%%%%%%%%%%%%%%%%%%%%%%%%%%%%%%%%%%%%%%
\section{a minimal model: $N=4$}
\label{sec:5}
%%%%%%%%%%%%%%%%%%%%%%%%%%%%%%%%%%%%%%%%%%%%%%%%%%%%%%%%%%%%%%%%%%%%%%
%%%%%%%%%%%%%%%%%%%%%%%%%%%%%%%%%%%%%%%%%%%%%%%%%%%%%%%%%%%%%%%%%%%%%%

In this section, we study a simple model for $N=4$ in detail.
According to Sec.~\ref{sec:3}, we obtain the finite part of the one-loop
potential as
\bea
&-&\frac{8}{2(4\pi)^{2}}\int_0^{\infty}
\frac{dt}{t}t^{-2}~
e^{-\overline{M}^2t}
\sum_{q_1=1}^{\infty}\sum_{q_2=-\infty}^{\infty}\left[
\cos(2q_1\varphi)
I_{2q_1}\left(f_1t\right)I_{2q_2-q_1}\left(f_2t\right)\right]\nn
&-&\frac{4}{2(4\pi)^{2}}\int_0^{\infty}
\frac{dt}{t}t^{-2}~
e^{-\overline{M}^2t}
\left[
I_{0}\left(f_1t\right)I_{0}\left(f_2t\right)-1-\frac{f_1^2+f_2^2}{4}t^2
\right]\nn
&-&\frac{8}{2(4\pi)^{2}}\int_0^{\infty}
\frac{dt}{t}t^{-2}~
e^{-\overline{M}^2t}
{\sum_{q_2=2}^{\infty}}
I_{0}\left(f_1t\right)I_{2q_2}\left(f_2t\right)\nn
&-&\frac{8}{2(4\pi)^{2}}\int_0^{\infty}
\frac{dt}{t}t^{-2}~
e^{-\overline{M}^2t}\left[
I_{0}\left(f_1t\right)I_{2}\left(f_2t\right)-
\frac{f_2^2}{8}t^2\right]\, ,
\label{sch2}
\eea
Since we should take $f_1^2+2f_2^2=\bar{f}^2$ (const.), 
we parameterize $f_r$ as
\be
f_1=\bar{f}\cos\theta\, ,\qquad
\sqrt{2}f_2=\bar{f}\sin\theta\, .
\ee

\begin{figure}[htb]
\centering
\mbox{\epsfbox{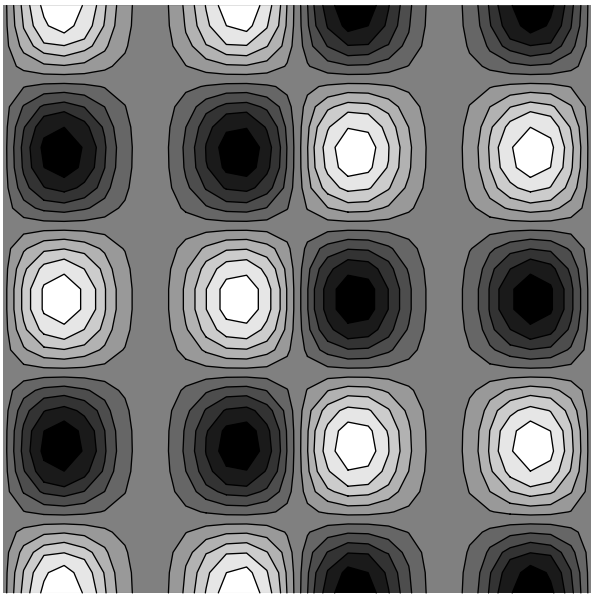}}\\
%\mbox{(a)}\\
%\mbox{\epsfbox{scalar2.eps}}\\
%\mbox{(b)}\\
\bigskip
\caption{%
Contour plot of the potential in the small $\bar{f}/\overline{M}^2$
limit.}
\label{fig1}
\end{figure}

\begin{figure}[htb]
\centering
\mbox{\epsfbox{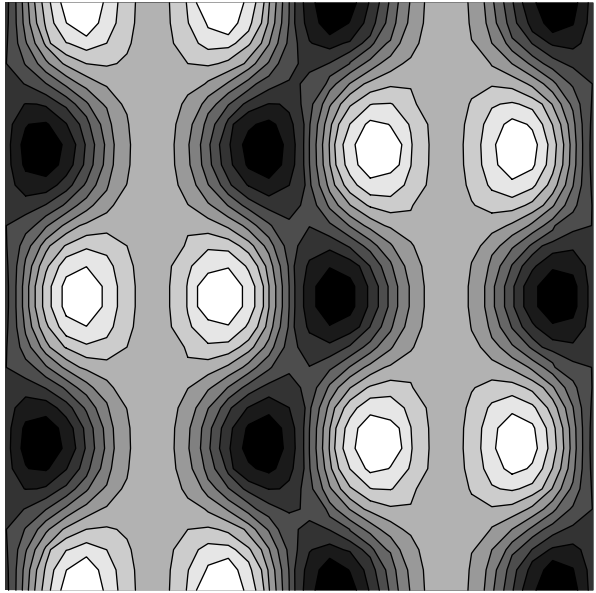}}\\
%\mbox{(a)}\\
%\mbox{\epsfbox{cp2.eps}}\\
%\mbox{(b)}\\
\bigskip
\caption{%
Contour plot of the potential for $\bar{f}/\overline{M}^2=0.5$.}
\label{fig2}
\end{figure}

At first sight of each term, the potential is expected to be of order
${\bar{f}^4}/{\overline{M}^4}$. However, using resummation
according to (\ref{formu}), we find
\bea
&-&\frac{8}{2(4\pi)^{2}}\int_0^{\infty}
\frac{dt}{t}t^{-2}~
e^{-\overline{M}^2t}
\sum_{\ell=1}^{\infty}\left[
\cos((4\ell-2)\varphi)
I_{4\ell-2}\left(\bar{f}\cos\theta
t\right)\sinh\left(\frac{\bar{f}}{\sqrt{2}}\sin\theta t\right)\right]\nn
&-&\frac{8}{2(4\pi)^{2}}\int_0^{\infty}
\frac{dt}{t}t^{-2}~
e^{-\overline{M}^2t}
\sum_{\ell=1}^{\infty}\left[
\cos(4\ell\varphi)
I_{4\ell}\left(\bar{f}\cos\theta
t\right)\cosh\left(\frac{\bar{f}}{\sqrt{2}}\sin\theta t\right)\right]\nn
&-&\frac{4}{2(4\pi)^{2}}\int_0^{\infty}
\frac{dt}{t}t^{-2}~
e^{-\overline{M}^2t}
\left[
I_{0}\left(\bar{f}\cos\theta t\right)\cosh
\left(\frac{\bar{f}}{\sqrt{2}}\sin\theta
t\right)-1-\frac{\bar{f}^2}{4}t^2
\right]\, ,
\label{sch3}
\eea
and the potential for $\bar{f}\ll\overline{M}^2$ is estimated as
\be
-\frac{1}{2\sqrt{2}(4\pi)^{2}}
\frac{\bar{f}^3}{\overline{M}^2}
\cos 2\varphi\sin\theta\cos^2\theta+O(\bar{f}^4/\overline{M}^4)
\, .
\ee
The structure of the potential in the small limit of
$\bar{f}/\overline{M}^2$ is shown in FIG.~\ref{fig1}.
Similarly the structure of the potential for finite
$\bar{f}/\overline{M}^2$ is shown in FIG.~\ref{fig2}.
In these figures, horizontal axes indicate $\theta$ while vertical ones
$\varphi$. Both variables are taken in the enlarged parameter region
$(-\pi,\pi)$.
Though the location of the minimum and maximum points are slightly
changed  according to the value of $\bar{f}/\overline{M}^2$,
number of extrema is unchanged.
Therefore a non-trivial expectation value selected by a potential
minimum is not very sensitive to the value $\bar{f}/\overline{M}^2$.

The (mass)${}^2$ eigenvalues $M^2_p$ associated with the potential
minimum are $M_1^2=M_2^2=M_3^2>M_4^2$ (and the permutations among them)
in the limit of
$\bar{f}\ll\overline{M}^2\rightarrow 0$. If we use fermion degrees
of freedom
instead of scalar fields, the (mass)${}^2$ eigenvalues $M^2_p$
associated with the potential minimum will be $M_1^2=M_2^2=M_3^2<M_4^2$
(and the permutations among them)
in the limit of
$\bar{f}\ll\overline{M}^2\rightarrow 0$.
The approximately degenerate masses except for one is expected to
be realized at the potential minimum in more general cases for $N>4$; to
see this, we estimate the integral expression for the potential by using
the asymptotic behavior of the modified Bessel function (for fixed
$\varphi_r$).

Interestingly enough, the mass of the scalars are expected to be
very small as $O(\bar{f}^2/\overline{M}^2)$ if
$\bar{f}\ll\overline{M}^2$, provided that the kinetic term is of order
of such as
$\bar{f}(\partial\theta)^2$.
%The small masses for scalars indicate the similar scenario for
%quintessence in the model of Hill and Leivobich~\cite{HL}.
Unfortunately,  since we do not know the origin of the kinetic term
at the present analysis, we cannot tell the precise order of the mass. 

%%%%%%%%%%%%%%%%%%%%%%%%%%%%%%%%%%%%%%%%%%%%%%%%%%%%%%%%%%%%%%%%%%%%%%
%%%FIGURES 1
%%%%%%%%%%%%%%%%%%%%%%%%%%%%%%%%%%%%%%%%%%%%%%%%%%%%%%%%%%%%%%%%%%%%%%
%%%%%%%%%%%%%%%%%%%%%%%%%%%%%%%%%%%%%%%%%%%%%%%%%%%%%%%%%%%%%%%%%%%%%%
%%%FIGURES 1
%%%%%%%%%%%%%%%%%%%%%%%%%%%%%%%%%%%%%%%%%%%%%%%%%%%%%%%%%%%%%%%%%%%%%%
%%%%%%%%%%%%%%%%%%%%%%%%%%%%%%%%%%%%%%%%%%%%%%%%%%%%%%%%%%%%%%%%%%%%%%
%%%FIGURES 2
%%%%%%%%%%%%%%%%%%%%%%%%%%%%%%%%%%%%%%%%%%%%%%%%%%%%%%%%%%%%%%%%%%%%%%
%%%%%%%%%%%%%%%%%%%%%%%%%%%%%%%%%%%%%%%%%%%%%%%%%%%%%%%%%%%%%%%%%%%%%%
%%%FIGURES 2
%%%%%%%%%%%%%%%%%%%%%%%%%%%%%%%%%%%%%%%%%%%%%%%%%%%%%%%%%%%%%%%%%%%%%%

Another interesting feature is the row of the minima in the case with
finite $\bar{f}/\overline{M}^2$ in FIG.~\ref{fig2}.
Some topological configurations which connect the minima along the
valley can be expected. Further study on the solitonic objects in
the model should be done.

%%%%%%%%%%%%%%%%%%%%%%%%%%%%%%%%%%%%%%%%%%%%%%%%%%%%%%%%%%%%%%%%%%%%%%
%%%%%%%%%%%%%%%%%%%%%%%%%%%%%%%%%%%%%%%%%%%%%%%%%%%%%%%%%%%%%%%%%%%%%%
\section{conclusion and discussion}
\label{sec:f}
%%%%%%%%%%%%%%%%%%%%%%%%%%%%%%%%%%%%%%%%%%%%%%%%%%%%%%%%%%%%%%%%%%%%%%
%%%%%%%%%%%%%%%%%%%%%%%%%%%%%%%%%%%%%%%%%%%%%%%%%%%%%%%%%%%%%%%%%%%%%%

In conclusion, we have shown that the one-loop finite  potential for
$N-2$ scalars is obtained from $N$ quantum fields.
Though the fact may have been already known,
we have explicitly found the $N-2$ degrees of freedom
in the parameterization of the mass spectrum of quantum fields.
To this end, we have utilized the expansion in terms of the modified
Bessel functions.

The location of the potential extrema is not so sensitive to the two
scales in the model. At the potential minimum, the mass eigenvalues of
$N$ quantum fields are expected to be almost degenerate except for one.
Therefore our model may provide us with a mechanism for spontaneous mass
splitting of several fields.

The application to the particle-theory model is expected.
We must make effort to clarify what symmetry enforces the mass matrix of
$N$ fields and the (probably gauged) kinetic term of $N-2$ scalars to be
the appropriate forms, and how symmetry breaking is triggered by the
expectation value of scalars when gauge symmetry is incorporated. For
this purpose, we have to take also diverse type of fields and their
quantum effects into account. In this paper, we have only treated the
scalar quantum field. We should consider the one-loop effect of fermions
and gauge bosons for more natural particle theory. On the other hand, the
higher-loop effects should be studied when interactions are introduced.
We will perform more analyses of the effective potential for general
models.

We can also cancel the scalar-independent divergent terms such as
(\ref{odd}) and (\ref{even}) by using fermionic quantum
fields as well as bosonic fields without supersymmetry. This possiblity
may shed light on new aspects of the cosmological constant problem
and inflation mechanism.

In any case, the cosmological implications of the model will be revealed
after analyzing more realistic models.
Nevertheless we anticipate that the light scalar degrees of freedom
becomes a candidate of dark matter or quintessence%
%, as in the model of Hill and Leivobich~\cite{HL}
. Moreover, it may be
interesting to study the finite temperature effect on the potential in
the hot early universe.
%; it will be discussed elsewhere~\cite{fw}.

%%%%%%%%%%%%%%%%%%%%%%%%%%%%%%%%%%%%%%%%%%%%%%

%%%%%%%%%%%%%%%%%%%%%%%%%%%%%%%%%%%%%%%%%%%%%%
\begin{acknowledgments}
We would like to thank 
 R.~Takakura 
for their valuable comments
and for the careful reading of the manuscript.
\end{acknowledgments}

%\newpage

%%%%%%%%%%%%%%%%%%%%%%%%%%%%%%%%%%%%%%%%%%%%%%%%%%%%%%%%%%%%%%%%%%%%%%
%%%References
%%%%%%%%%%%%%%%%%%%%%%%%%%%%%%%%%%%%%%%%%%%%%%%%%%%%%%%%%%%%%%%%%%%%%%

%%%%%%%%%%%%%%%%%%%%%%%%%%%%%%%%%%%%%%%%%%%%%%%%%%%%%%%%%%%%%%%%%%%%%%
\end{document}